\documentclass[conference]{IEEEtran}
\IEEEoverridecommandlockouts

\usepackage{amsmath}
\usepackage{graphicx}
\usepackage{algorithm}
\usepackage{algorithmic}
\usepackage{amsfonts}
\usepackage{hyperref}
\usepackage{caption}
\usepackage{subcaption}
\usepackage{tikz}
\usepackage{footnote}
\usetikzlibrary{positioning}
\usepackage{CJKutf8}

\usepackage{cite}
\usepackage{amsmath,amssymb,amsfonts}
\usepackage{algorithmic}
\usepackage{graphicx}
\usepackage{textcomp}
\usepackage{xcolor}
\def\BibTeX{{\rm B\kern-.05em{\sc i\kern-.025em b}\kern-.08em
    T\kern-.1667em\lower.7ex\hbox{E}\kern-.125emX}}

\begin{document}
\begin{CJK}{UTF8}{gbsn}
\title{A Seesaw Model Attack Algorithm for Distributed Learning}


\author{Kun Yang$^{\ddag}$， Tianyi Luo$^{\ddag}$, Yanjie Dong$^\ddag$, and Aohan Li$^{\natural}$\\
	\mbox{$^\dag$Artificial Intelligence Research Institute, Shenzhen MSU-BIT University, Shenzhen 518172, China}\\
	\mbox{$^\natural$Graduate School of Informatics and Engineering, The University of Electro-Communications, Tokyo, Japan.}
	\thanks{This work was supported by the National Nature Science Foundation of China (62102266).}
}

\maketitle
\begin{abstract}
We investigate the Byzantine attack problem within the context of model training in distributed learning systems. While ensuring the convergence of current model training processes, common solvers (e.g. SGD, Adam, RMSProp, etc.) can be easily compromised by malicious nodes in these systems. Consequently, the training process may either converge slowly or even diverge.To develop effective secure distributed learning solvers, it is crucial to first examine attack methods to assess the robustness of these solvers. In this work, we contribute to the design of attack strategies by initially highlighting the limitations of finite-norm attacks. We then introduce the seesaw attack, which has been demonstrated to be more effective than the finite-norm attack. Through numerical experiments, we evaluate the efficacy of the seesaw attack across various gradient aggregation rules. 
\end{abstract}


\section{Introduction}
Federated learning refers to the utilization of multiple nodes and a server to work collaboratively for model training over large-scale datasets \cite{Kairouz2021}. 
Due to the data explosion and the ever-increasing complexity of machine learning models, the model training process has seen a paradigm shift from traditional single-machine training to multiple-machine federated training \cite{Li2020}. 
By expecting the multiple-machine cluster to process more data and handle more complex models, federated learning has become a promising research direction \cite{Gao2021}. 
Moreover, federated learning is a large-scale learning method that significantly and effectively improves the speed of model training in machine learning and deep learning tasks due to the fact that in federated learning, data is distributed across multiple client devices for parallel processing \cite{Xie2019}. 
Given the limited computational resources of a single node, leveraging computational resources across multiple machines allows the training of large models with massive parameters \cite{Feng2020}. 
As large models advance, distributed learning is increasingly used.

In federated learning, each participating node trains locally based on the received dataset \cite{Konecny2016FederatedLS, yd2019} and submits the gradients to the central server for aggregation \cite{Liu2019}. 
The aggregated gradient updates the parameters of the global model. 
However, due to aggregation ensuring the confidentiality of local model training on each participating node \cite{McMahan2017, ydarxiv}, the central server cannot detect whether any participating node's contribution to the global model is abnormal.

In federated learning, any compromised node can use strategies to replace or alter the gradients sent to the central server, thereby interfering with or disrupting the central server's model training. For instance, Byzantine nodes can generate random gradients or calculate a gradient opposite to the correct update direction to prevent the model from updating correctly, leading to poor accuracy \cite{Liu2020}.

To address the Byzantine problem, several gradient aggregation rules (GARs) have been proposed. One classical GAR is the Krum aggregation \cite{Blanchard2017MachineLW,9}, which selects gradients based on their distance relationships to avoid including malicious gradients in the final model update. Whilst, El Mahdi El et al. \cite{ElMhamdi2018TheHV} proposed a norm attack method that can bypass Krum aggregation by exploiting the use of norms as distance measures. Moreover, recent work has shown vulnerabilities in distributed learning systems that could be exploited by Byzantine nodes \cite{Mhamdi2020, Shen2020}.

\subsection{Contributions}
In this paper, we analyze the principle of the finite-paradigm attack on Krum aggregation, from which we propose the seesaw attack. Furthermore, we experimentally prove that the seesaw attack is more effective than the finite-paradigm attack in destroying Krum aggregation. The specific contributions are as follows:
\begin{itemize}
\item
We analyzed the nature of limited norm attacks \cite{11}. Essentially, limited norm attacks exploit the ambiguity of using norms as a distance metric in the Krum aggregation algorithm. A fixed distance might result from a significant contribution in one dimension or slight contributions across multiple dimensions. Thus, changing the distance metric or detecting single-dimensional differences can effectively counteract the ambiguity brought by norms, defending against norm attacks.
\item
Based on the analysis of the nature and limitations of norm attacks, we proposed another attack method, seesaw attack, for the Krum aggregation strategy, which continues the idea of directional deviation attacks from norm attacks.
\item
We aim to bolster the robustness of the Krum aggregation algorithm by integrating hierarchical clustering. It categorizes gradients from nodes into five groups, computes category averages as representatives, mitigating malicious node influence.
\item
We compared the effectiveness of our designed seesaw attack and norm attacks under different aggregation algorithms on the MNIST dataset. Experimental results show that the seesaw attack has a superior disruptive effect on the Krum aggregation algorithm compared to norm attacks.
\end{itemize}

\section{Federated Learning Gradient Aggregation}
\subsection{Federated Learning Model with Byzantine Nodes}
We use the distributed data parallel (DDP) model to improve the efficiency of distributed training. The distributed model allows parallelization of the training process across multiple GPUs or machines, accelerating model training. Each node computes gradients locally and synchronizes these gradients with others, resulting in a gradient vector for updating model parameters. However, if some nodes are compromised, the malicious gradients can significantly disrupt the model's training.

Under this kind of background, attackers are defined as entities controlling $f$ nodes. By collecting gradients from the other $n-f$ non-Byzantine nodes, attackers can manipulate the $f$ nodes to send malicious gradients, influencing the final update direction and model performance.The adversary can understand the system state at any given time,including:
\begin{itemize}
\item
The complete state of the master system (data and code)
\item
The complete state of each node (data and code)
\item
Any data exchanged through communication channels
\end{itemize}

Thus, the attacker can leverage its knowledge of the master state and the submitted gradients to construct effective attacks. However, the adversary is not omnipotent; it cannot directly alter the system state, impersonate other nodes, or delay communication. The attacker can only submit gradients through controlled nodes.

\subsection{Gradient Aggregation Rules (GAR)}

In distributed learning frameworks, a specific policy rule is usually needed to aggregate the gradients computed at each node, and common gradient aggregation methods are:

\subsubsection{Weighted Average}
The simplest aggregation technique is FedSGD \cite{McMahan2017}, which performs a simple weighted mean aggregation of the gradients based on the number of data samples each client holds.

\subsubsection{Trimmed Mean}
Trimmed mean removes a portion of the highest and lowest values before averaging to reduce the impact of malicious nodes reporting extreme values.

Assume we have a set of values reported by nodes $\{x_1, x_2, \ldots, x_n\}$, among which there are $f$ malicious nodes. The steps for calculating the trimmed mean are as follows:

\begin{enumerate}
  \item Sort the values to obtain the ordered set $\{x_{(1)}, x_{(2)}, \ldots, x_{(n)}\}$.
  \item Remove the smallest $f$ values and the largest $f$ values, leaving the set $\{x_{(f+1)}, x_{(f+2)}, \ldots, x_{(n-f)}\}$.
  \item Calculate the average of the remaining values, which is:
\end{enumerate}

\begin{equation}
\text{Trimmed mean} = \frac{1}{n-2f} \sum_{i=f+1}^{n-f} x_{(i)}
\end{equation}

\subsubsection{Median}
The median method selects the median value for aggregation, providing robust resistance to noise and malicious nodes.

Assume we have a set of values reported by nodes $\{x_1, x_2, \ldots, x_n\}$. The steps for calculating the median are as follows:

\begin{enumerate}
  \item Sort the values to obtain the ordered set $\{x_{(1)}, x_{(2)}, \ldots, x_{(n)}\}$.
  \item If $n$ is odd, the median is the middle value of the sorted list, which is:
  \begin{equation}
  \text{Median} = x_{\left(\frac{n+1}{2}\right)}
  \end{equation}
  \item
  If $n$ is even, the median is the average of the two middle values of the sorted list, which is:
  \begin{equation}
  \text{Median} = \frac{x_{\left(\frac{n}{2}\right)} + x_{\left(\frac{n}{2} + 1\right)}}{2}
  \end{equation}
\end{enumerate}

\subsubsection{Krum}
Krum aggregation algorithm selects a gradient with the smallest total distance to other gradients, excluding those that might be generated by malicious nodes. The steps include:

\begin{enumerate}
  \item Calculate the Euclidean distance:
  For each model update $v_i$ and $v_j$, calculate the Euclidean distance between them:
  \begin{equation}
  d(v_i, v_j) = \| v_i - v_j \|_2
  \end{equation}
  where $\| \cdot \|_2$ denotes the Euclidean norm.

  \item Calculate the total distance:
  For each update $v_i$, calculate the sum of its distances to the $n - f - 2$ closest other updates:
  \begin{equation}
  S(i) = \sum_{v_j \in N(i)} d(v_i, v_j)
  \end{equation}
  where $N(i)$ denotes the set of the $n - f - 2$ closest updates to $v_i$.

  \item Choose the update with the smallest total distance:
  Find the update $v_i$ that minimizes $S(i)$:
  \begin{equation}
  \hat{v} = \arg \min_{v_i} S(i)
  \end{equation}
  The selected update $\hat{v}$ is used to update the global model.
\end{enumerate}

\subsubsection{FABA}
Federated Averaging with Byzantine-resilient Aggregation(FABA) is a context-specific method that iteratively eliminates the gradients that deviate the most from the norm, with the aim of mitigating the effects of Byzantine faults in distributed machine learning systems. These Byzantine faults refer to malicious nodes or failures that can disrupt the normal operation of the system by sending incorrect or misleading information.

FABA iteratively removes the most deviating gradients to counter Byzantine faults.

\subsubsection{Geometric Median with Cosine Similarity}
This method measures the cosine similarity between gradient vectors to select the most representative gradients, excluding those significantly different.

\section{Directional Deviation Attack}
\subsection{Analysis of Directional Deviation Attack}

In most Gradient Aggregation Rules (GARs), gradients are aggregated by computing the distances between a given vector and all other vectors, with the goal of selecting gradients that are close in distance to contribute to the aggregation process. This method aims to alleviate the impact of gradients generated maliciously by Byzantine nodes, which may significantly deviate from or even oppose the correct update direction, thereby preventing them from distorting the final aggregation outcome.

Upon rational analysis, the impact of Byzantine nodes on gradient aggregation can be categorized into two main aspects. First, there is the Euclidean distance between malicious gradients produced by Byzantine nodes and normal gradients. Metrics such as the $\ell_2$-norm can effectively identify these deviations and mitigate the influence of gradients with substantial discrepancies on the final aggregation process\cite{Schaeffer2013}.

Second, the alignment of gradient vectors plays a crucial role. If Byzantine nodes generate a significant number of malicious vectors that are closely aligned or even overlap in proximity, distance-based aggregation algorithms are likely to select these vectors as the final gradient direction for model parameter updates. When these malicious gradients diverge from the true update direction, it can result in diminished accuracy of the trained model\cite{Salton2012}.

A classic example is the norm-based attack method proposed by EI Mahdi EI Mhamdi et al., where each Byzantine node first collects gradients from all non-Byzantine nodes, computes the average of these normal gradients, and then makes small manipulations on this average gradient vector in single dimensions. Consequently, all gradients generated by Byzantine nodes become identical, with a Euclidean distance of 0 between them. Furthermore, these forged gradients from Byzantine nodes are not far from normal gradients in distance, leading distance-based aggregation rules like Krum to select malicious gradient vectors as the final update direction. Thus, attackers can control the direction of gradient updates within a certain range\cite{Hardt2017}.

\subsection{Seesaw Attack}

Compared to the norm-based attack proposed by EI Mahdi EI Mhamdi et al., we no longer select the average value of gradients generated by normal nodes as the reference gradient for generating malicious attack gradients. Instead, we choose the median gradient among gradients generated by normal nodes as the reference gradient to generate malicious gradients.

These malicious nodes are all omniscient and omnipotent. They possess comprehensive knowledge of the model's structure and parameters, as well as the ability to manipulate the gradients with precision. This allows them to launch sophisticated attacks, such as creating gradients that appear benign while subtly steering the model towards a biased or compromised state. Their omniscience enables them to anticipate defensive mechanisms, while their omnipotence ensures they can adapt their strategies to bypass these defenses effectively.

Taking Krum as an example of a GAR, the Krum aggregation algorithm first selects each gradient vector $V_i$ and calculates the $\ell_2$-norm distance between $V_i$ and every other gradient vector $V_j$ (excluding itself). For explanatory convenience, in this paper, we refer to $V_i$ as the "principal vector." Then, the algorithm selects the $n-f-2$ closest gradient vectors to the principal vector based on the $\ell_2$-norm distance and calculates a score (where score equals the sum of $\ell_2$-norm distances between the principal vector and the other $n-f-2$ closest gradient vectors).

By adopting the seesaw attack method, it is equivalent to having $f+1$ nodes with gradient values that are nearly identical. Moreover, these gradient vectors are selected based on the median of gradient vectors generated by normal nodes. Therefore, according to the Krum algorithm for selecting update gradients, this median gradient will inevitably be chosen as the update gradient. If the model updates according to this gradient direction, it primarily benefits one normal node, while updates for other nodes are biased. Additionally, since each card is allocated datasets that have no intersection with each other, this attack strategy will effectively result in only one card's dataset playing a role in the model training during a training round\cite{Wang2020}.

\section{Hierarchical Clustering Krum}
\subsection{Hierarchical Clustering for Enhanced Krum Aggregation}  
We endeavor to bolster the robustness of the Krum aggregation algorithm by incorporating hierarchical clustering. This approach categorizes the gradients gathered from multiple nodes into five distinct clusters and computes the average gradient of each cluster as its representative. This preprocessing step serves to mitigate the adverse effects of malicious nodes on the overall model performance.  

Hierarchical clustering Krum offers several advantages: enhanced robustness, as it diminishes the influence of malicious nodes on the model by segregating gradients into multiple clusters; cluster representativeness, where the average gradient of each cluster is adopted as its representative, resulting in a more stable aggregation outcome and reducing the impact of outliers; improved aggregation efficiency, since only a reduced set of cluster representatives are processed instead of gradients from individual nodes, enhancing computational efficiency; and reduced computational complexity, as calculations of distances and neighbors are confined to cluster representatives, leading to decreased computational complexity. However, there are potential clustering errors, where the algorithm may inadvertently group gradients from malicious nodes with those from benign nodes, compromising the accuracy of the aggregation result\cite{Jain2010}.

Whereas, additional computational overhead is introduced, as the clustering process itself necessitates more computational time and resources, particularly when dealing with large-scale datasets. Bias in cluster representatives may occur, as the average gradients may not accurately mirror the true state of all cluster members, introducing deviations in the aggregation outcome. Increased complexity is another issue, as the inclusion of the clustering step heightens the algorithm's complexity, making implementation and debugging more challenging. However, the motivation of this method is to "throw a spart to catch a whale" and trigger more strategies to defense the directional deviation attack\cite{Huang2010, Yang2020}.


\section{Experiments}

\subsection{Experiment Setup}
\subsubsection{Hardware Environment}
The experiments were conducted on a server equipped with four NVIDIA GeForce RTX 4090 GPUs. The server runs on Ubuntu 20.04 LTS, with an Intel Xeon processor and 128GB of memory. To ensure reproducibility, all experimental steps were carried out in this hardware environment.

\subsubsection{Experimental Parameter Design}

We assume that the distributed system consists of $n$ nodes, with $f$ of them being Byzantine nodes. In our experiment, we set $n$ to 7 and $f$ to 4, satisfying the condition $n > 2f + 3$, which ensures the system's robustness and integrity in the presence of Byzantine failures. This setup guarantees that the total number of nodes exceeds twice the number of Byzantine nodes plus three, a critical requirement for Byzantine fault tolerance.

For the training process, we utilized the MNIST dataset, a widely recognized benchmark for image classification. The model was trained for 10 epochs using the Adam optimizer, with the learning rate set to 0.001. 

\subsubsection{Network Architecture Design}
\begin{figure}[ht]
    \centering
    \begin{tikzpicture}[node distance=0.2cm and 0.2cm, 
                        every node/.style={draw, minimum height=0.5cm, minimum width=0.3cm, align=center}]
        \node (input) {Input (1x28x28)};
        \node [below=of input] (conv1) {Conv2d (1, 32, 3x3)};
        \node [below=of conv1] (relu1) {ReLU};
        \node [below=of relu1] (conv2) {Conv2d (32, 64, 3x3)};
        \node [below=of conv2] (relu2) {ReLU};
        \node [below=of relu2] (pool) {Max Pooling (2x2)};
        \node [below=of pool] (dropout1) {Dropout (0.25)};
        \node [below=of dropout1] (flatten) {Flatten};
        \node [below=of flatten] (linear1) {Linear (9216 to 128)};
        \node [below=of linear1] (relu3) {ReLU};
        \node [below=of relu3] (dropout2) {Dropout (0.5)};
        \node [below=of dropout2] (linear2) {Linear (128 to 10)};
        \node [below=of linear2] (logsoftmax) {Log Softmax};
        \node [below=of logsoftmax] (output) {Output (10)};
        
        \draw[->] (input) -- (conv1);
        \draw[->] (conv1) -- (relu1);
        \draw[->] (relu1) -- (conv2);
        \draw[->] (conv2) -- (relu2);
        \draw[->] (relu2) -- (pool);
        \draw[->] (pool) -- (dropout1);
        \draw[->] (dropout1) -- (flatten);
        \draw[->] (flatten) -- (linear1);
        \draw[->] (linear1) -- (relu3);
        \draw[->] (relu3) -- (dropout2);
        \draw[->] (dropout2) -- (linear2);
        \draw[->] (linear2) -- (logsoftmax);
        \draw[->] (logsoftmax) -- (output);
    \end{tikzpicture}
    \caption{Neural Network Architecture.}
    \label{fig:nn_architecture}
\end{figure}
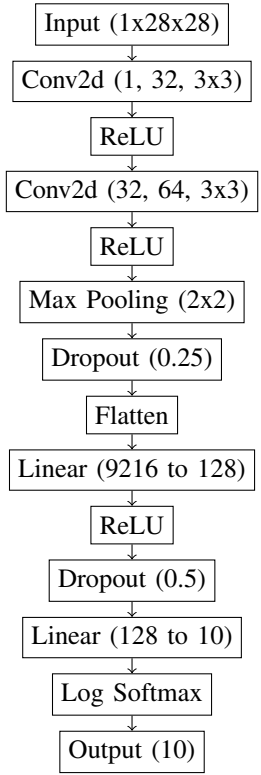
In Fig.~\ref{fig:nn_architecture}, the neural network's input and output, as well as special layers, are briefly described. The input layer receives an image of size 1x28x28, and the output layer is a Log Softmax layer with 10 classes. Special layers include two convolutional layers, each followed by a ReLU activation function, a max-pooling layer, a Dropout layer to prevent overfitting, a Flatten layer to flatten the features, and finally, two fully connected layers, with the last being the Log Softmax output.
\subsubsection{Dataset Partitioning}
The dataset is divided into 11 parts, each trained on a different GPU. Byzantine nodes collect gradients from all good nodes, calculate the geometric median, and generate malicious gradients with small perturbations. Krum aggregates the gradients by selecting $n-f-2$ nearest vectors and computes the central gradient.

To test the model's generalization, we split each GPU's dataset into 20\% test and 80\% training sets.
\subsection{Effectiveness of Seesaw Algorithm}
\begin{figure}[htbp]
    \centering
    \includegraphics[width=0.4\textwidth]{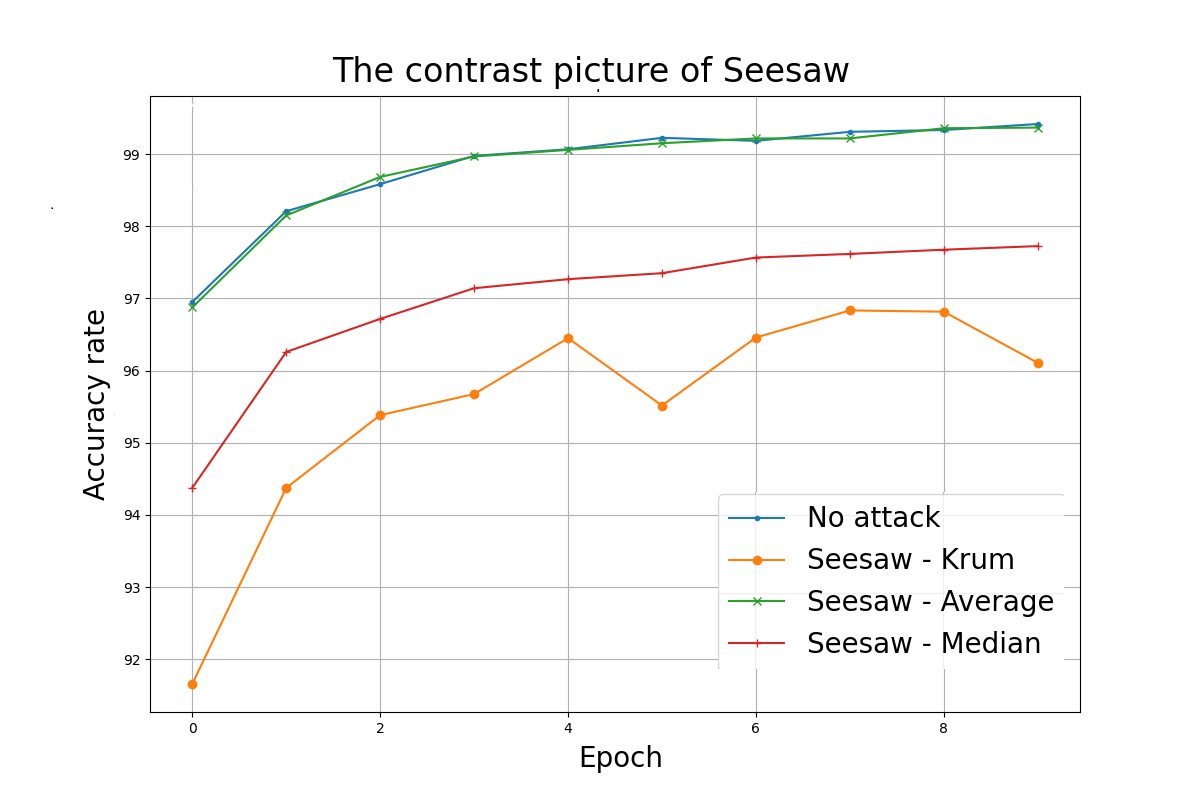}
    \caption{Comparison of the defense effects of different aggregation strategies under seesaw attacks}
    \label{fig:my_label2}
\end{figure}
In Fig.~\ref{fig:my_label2}, the Seesaw algorithm, compared to the non-attacked situation, resulted in an accuracy loss of 0.1\% with mean aggregation, 3.5\% with Krum aggregation, and 1.7\% with median aggregation, doubling the attack degree. The Seesaw algorithm caused the most disruption with Krum aggregation and the least with mean aggregation. The likely reason for this is the higher proportion of Byzantine nodes participating in the aggregation when using Seesaw attack in Krum aggregation, compared to a lower proportion in mean aggregation, thus making the Seesaw attack less effective under mean aggregation.

The accuracy achieved through mean aggregation without any attack was 99.5\%. When the Seesaw algorithm was applied, for Mean Aggregation, the accuracy was reduced by 0.1\%, dropping to 99.4\%. This minimal disruption is due to the mean aggregation method averaging across all nodes, thus diluting the impact of individual node attacks on the overall aggregation result.
For Krum Aggregation, the accuracy was reduced by 3.5\%, dropping to 97.0\%. The highest disruption is because Krum aggregation selects a single aggregation node, and the Seesaw algorithm's targeted attack on this node significantly reduces accuracy.
For Median Aggregation, the accuracy was reduced by 1.7\%, dropping to 97.8\%. The considerable disruption is because median aggregation selects the median node, and the Seesaw algorithm attacks nodes close to the median to influence the aggregation result, though not as significantly as with Krum aggregation.

\begin{figure}[htbp]
    \centering
    \includegraphics[width=0.4\textwidth]{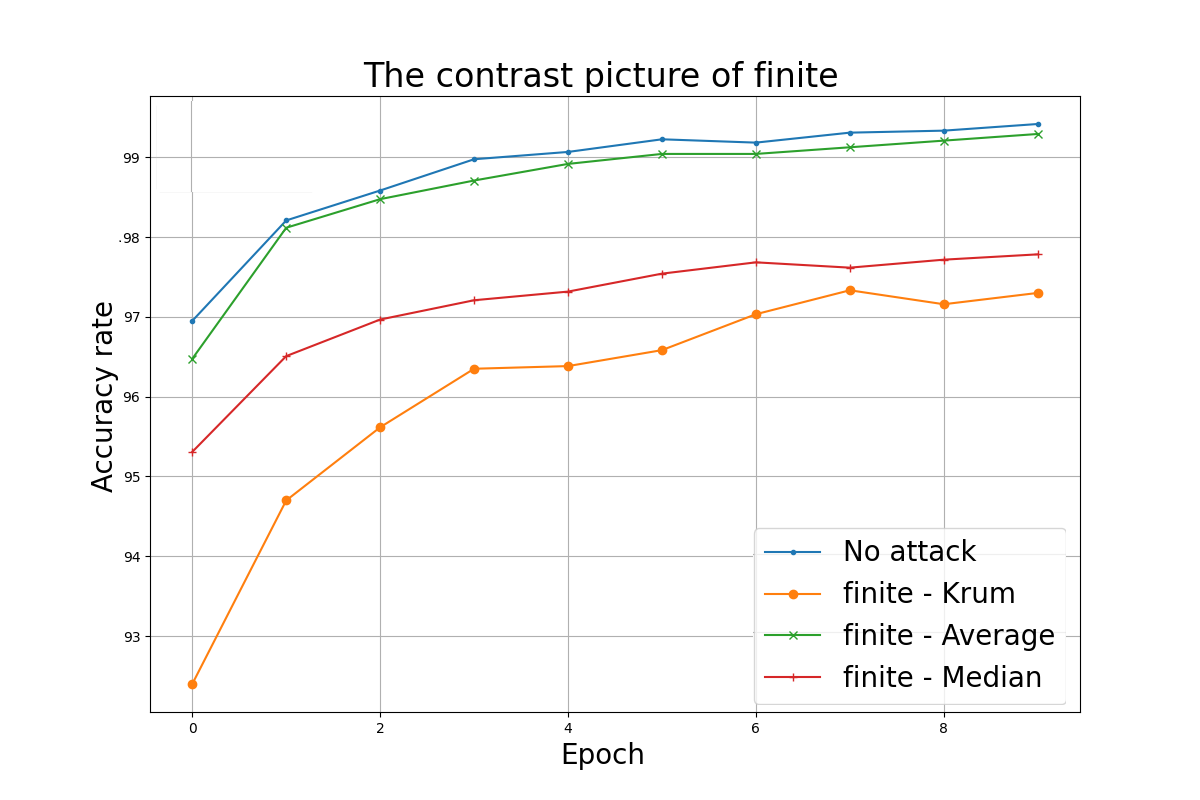}
    \caption{Comparison of the defense effects of different aggregation strategies under limited norm attacks}
    \label{fig:my_label3}
\end{figure}
In Fig.~\ref{fig:my_label3}, we discovered that the same characteristics also emerged in the horizontal control group of attacks with finite norm.

\begin{figure}[htbp]
    \centering
    \includegraphics[width=0.4\textwidth]{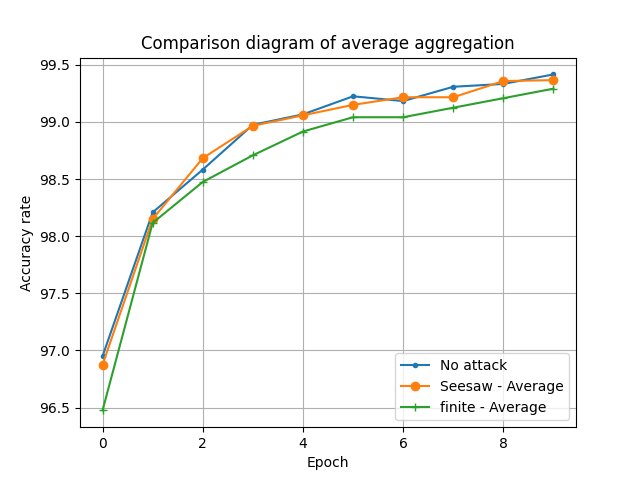}
    \caption{Comparison of defense effects under different attacks with average aggregation. The severity of the Seesaw attack is slightly lower than that of the limited norm attack.}
    \label{fig:my_label4}
\end{figure}

\begin{figure}[htbp]
    \centering
    \includegraphics[width=0.4\textwidth]{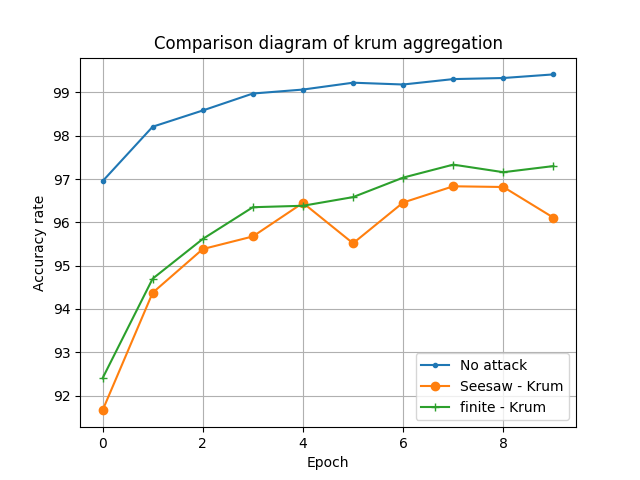}
    \caption{Comparison of defense effects under different attacks with Krum aggregation. The severity of the Seesaw attack is significantly higher than that of the limited norm attack}
    \label{fig:my_label5}
\end{figure}

\begin{figure}[htbp]
    \centering
    \includegraphics[width=0.4\textwidth]{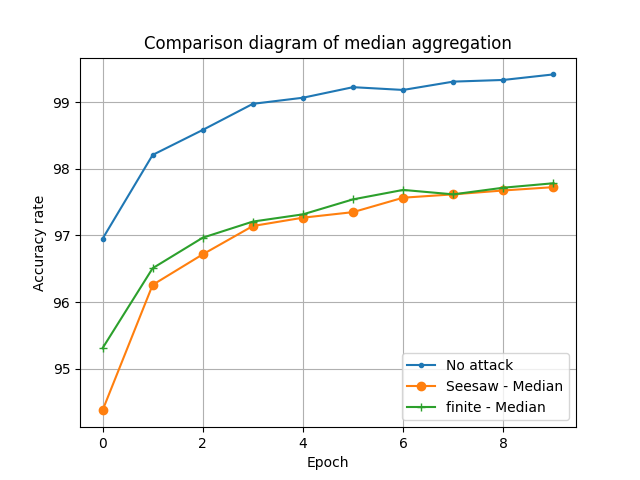}
    \caption{Comparison of defense effects under different attacks with geometric median aggregation. The severity of the Seesaw and limited norm attacks is similar}
    \label{fig:my_label6}
\end{figure}

We also compared the defense effects of different aggregation strategies under various attacks (Fig.~\ref{fig:my_label4}, Fig.~\ref{fig:my_label5}, Fig.~\ref{fig:my_label6}) and summarized the results. In summary, there are significant differences in the defense effectiveness of different aggregation methods. Under average aggregation, the defense effect is relatively stable and performs well against various attacks, particularly against the Seesaw attack. However, under the Krum aggregation strategy, the system is more sensitive to the Seesaw attack, showing poorer defense capabilities. On the other hand, the geometric median aggregation method shows a more balanced defense capability, handling both Seesaw and limited norm attacks with similar effectiveness.

The reason behind this lies in the geometric median's ability to minimize the influence of outliers by focusing on the median point, rather than the mean. This approach makes it harder for attackers to manipulate the overall gradient, as the geometric median is less sensitive to extreme values. Consequently, it offers a robust defense against both types of attacks by effectively neutralizing the impact of malicious gradients.

\subsection{Effectiveness of Hierarchical Clustering Krum}  
\begin{figure}[htbp]
    \centering
    \includegraphics[width=0.4\textwidth]{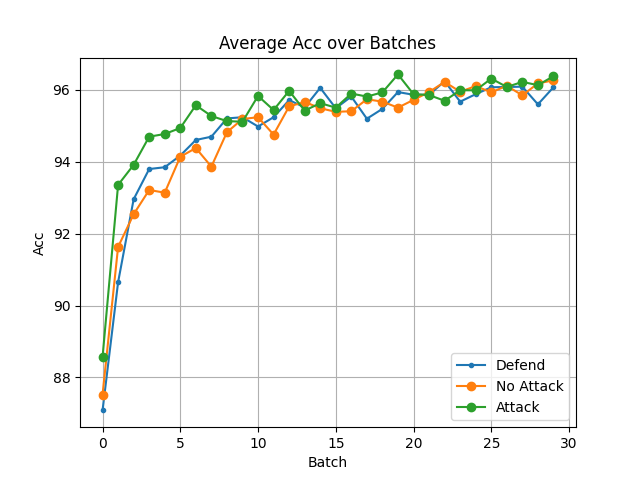}
    \caption{The influence curve of attack and defense on accuracy under hierarchical clustering}
    \label{fig:my_label7}
\end{figure}

Figure~\ref{fig:my_label7} illustrates the influence curve of attack and defense on accuracy under hierarchical clustering. Below is a detailed description of the different colored curves.
  
The accuracy of attack with defend initially lags in the earlier batches but gradually escalates and stabilizes, ultimately approaching 96\%. This trend signifies that hierarchical clustering Krum possesses the capability to progressively recover under attack and sustain high accuracy in the later phases of training.  
   
The accuracy of no attck consistently ascends across all batches, stabilizing towards the conclusion of training, and ultimately attains close to 96\%. This observation underscores the stability of hierarchical clustering Krum in training the model to high accuracy levels in the absence of any attack.  
    
The accuracy of attack without defend swiftly rises in the initial batches, peaking in the mid-stages of training, albeit with slight fluctuations in the subsequent stages, ultimately converging near 96\%. This pattern implies that average aggregation, when subjected to a finite norm attack, exhibits notable variability in accuracy yet generally attains high accuracy levels.

The primary reason why the averaging aggregation method can defend against finite norm attacks lies in its inherent robustness. Finite norm attacks manipulate the gradients of some nodes to cancel out their contributions in the overall gradient computation, thereby attempting to influence the final model update. However, the averaging aggregation method calculates the overall gradient by taking the arithmetic mean of all node gradients. Since malicious nodes usually constitute a minority, their manipulated gradients are diluted in the averaging process. As a result, the overall outcome still reflects the correct contributions of the majority of nodes, weakening the impact of the malicious gradients and effectively defending against such attacks.

This suggests that hierarchical clustering Krum defense and average aggregation are comparable in defending against limited norm attacks.

\subsection{Further Analysis Based on Experimental Results}

Based on the experimental results, the Seesaw attack has the most significant disruption on Krum aggregation. This is because the Krum aggregation method is designed to select the gradient that is closest to the other nodes as the update value. However, the Seesaw attack uploads multiple similar malicious gradients, causing these malicious nodes to be mistakenly selected by the Krum algorithm, resulting in more severe errors. This mechanism exploits Krum's vulnerability to small differences between malicious nodes, especially when multiple malicious nodes cooperate, making Krum more likely to select the gradient from a malicious node.

The Krum aggregation formula is:

\[
Krum(g) = \arg\min_{g_i} \sum_{j \in N(i)} \| g_i - g_j \|^2
\]

where \( N(i) \) represents the set of nearest neighbors to node \( i \). Malicious nodes upload similar but deviant gradients \( g_{\text{malicious}} = -c \cdot g_{\text{benign}} \), and by making the distances between each other small, Krum aggregation may choose one of these malicious gradients. This leads to significant deviation in the model update direction, as confirmed by experimental results.

On the other hand, mean aggregation is less disrupted by the Seesaw attack. This is because the mean aggregation method averages the gradients from all nodes. Although malicious nodes can upload amplified reverse gradients, the influence of the malicious gradients is diluted by the benign nodes, which are in the majority. Even if the malicious nodes upload amplified reverse gradients with a large factor \( c \), the mean aggregation result remains closer to the direction of the benign nodes. This mechanism shows that mean aggregation has stronger resilience to attacks, effectively mitigating the influence of malicious nodes.

The mean aggregation formula is:

\[
g_{\text{avg}} = \frac{1}{n} \left( (n-m) g_{\text{benign}} + m (-c \cdot g_{\text{benign}}) \right)
\]

where \( m \) is the number of malicious nodes, and \( c \) is the amplification factor. If the number of malicious nodes is small, or \( c \) is not too large, the mean aggregation effectively cancels out the malicious gradients. Even if the malicious nodes upload amplified gradients, experimental results show that the interference with the aggregation result is relatively small.

\subsection{Conclusion}

In conclusion, the Seesaw attack has the most significant effect under Krum aggregation, while the interference with mean aggregation is minimal. Through mathematical derivation and experimental analysis, we observe that the Seesaw attack exploits the vulnerabilities of different aggregation rules by using reverse amplified gradients. Krum aggregation's vulnerability lies in its sensitivity to small differences between malicious nodes, which can cause it to select an incorrect gradient. In contrast, mean aggregation is better at canceling out the influence of malicious nodes. By further optimizing these aggregation methods, the robustness of distributed learning systems can be improved.

\section{Concluding Remarks}
In this paper, we have initially investigated the finite-paradigm attack and figure out that its core mechanism involves exploiting the ambiguity of the distance metric in Krum aggregation algorithm to implement a directional deviation attack on a single dimension. This method achieves the attack by falsifying data exclusively on the first dimension.

Subsequently, according to the concept of a directional bias attack, we have proposed a more effective method to interfere with the gradient aggregation process. Specifically, we selected the median gradient generated by normal nodes as the benchmark, and all Byzantine nodes were engineered to produce gradients resembling this benchmark. Experimental results have confirmed that the seesaw attack is particularly effective in undermining the Krum aggregation algorithm.

The findings indicate that the seesaw attack reduces the accuracy by 3.5\% under the Krum aggregation algorithm and also exerts some level of interference under the mean and median aggregation algorithms. Compared to the finite-paradigm attack, the seesaw attack has proven to be more disruptive to model training.

\end{CJK}
\end{document}